\documentclass[sigconf]{acmart}
\AtBeginDocument{%
  }

\usepackage{algorithm}
\usepackage{algpseudocode}
\usepackage{amsmath}
\usepackage{amsthm}
\usepackage{multirow}
\usepackage[most]{tcolorbox}
\usepackage{latexsym}
\usepackage{pifont}

\AtBeginDocument{%
  }

\setcopyright{acmlicensed}
\copyrightyear{2018}
\acmYear{2018}
\acmDOI{XXXXXXX.XXXXXXX}
\acmConference[SIGKDD '25 Agent4IR Workshop]{Make sure to enter the correct
  conference title from your rights confirmation email}
  {August 3-7th 2025}{Toronto, ON, Canada}
  
\begin{document}

\title[AMD: A New Query Expansion Approach via Agent-Mediated Dialogic Inquiry]{A New Query Expansion Approach for Enhancing Information Retrieval via Agent-Mediated Dialogic Inquiry}

\author{Wonduk Seo}
\affiliation{%
  \institution{Enhans}
  \country{Seoul, South Korea}
}
\email{wonduk@enhans.ai}

\author{Hyunjin An}
\affiliation{%
  \institution{Enhans}
  \city{Seoul}
  \country{South Korea}}
\email{hyunjin@enhans.ai}

\author{Seunghyun Lee}
\affiliation{%
  \institution{Enhans}
  \city{Seoul}
  \country{South Korea}}
\email{seunghyun@enhans.ai}

\renewcommand{\shortauthors}{Seo et al.}

\begin{abstract}
Query expansion is widely used in Information Retrieval (IR) to improve search outcomes by supplementing initial queries with richer information. While recent Large Language Model (LLM) based methods generate pseudo-relevant content and expanded terms via multiple prompts, they often yield homogeneous, narrow expansions that lack the diverse context needed to retrieve relevant information. In this paper, we propose \textbf{AMD}: a new \textbf{\underline{A}}gent-\textbf{\underline{M}}ediated \textbf{\underline{D}}ialogic Framework that engages in a dialogic inquiry involving three specialized roles: (1) a \emph{Socratic Questioning Agent} reformulates the initial query into three sub-questions, with each question inspired by a specific Socratic questioning dimension, including clarification, assumption probing, and implication probing, (2) a \emph{Dialogic Answering Agent} generates pseudo-answers, enriching the query representation with multiple perspectives aligned to the user's intent, and (3) a \emph{Reflective Feedback Agent} evaluates and refines these pseudo-answers, ensuring that only the most relevant and informative content is retained. By leveraging a multi-agent process, \textbf{AMD} effectively crafts richer query representations through inquiry and feedback refinement. Extensive experiments on benchmarks including BEIR and TREC demonstrate that our framework outperforms previous methods, offering a robust solution for retrieval tasks.

\end{abstract}


\begin{teaserfigure}
    \includegraphics[width=\textwidth]{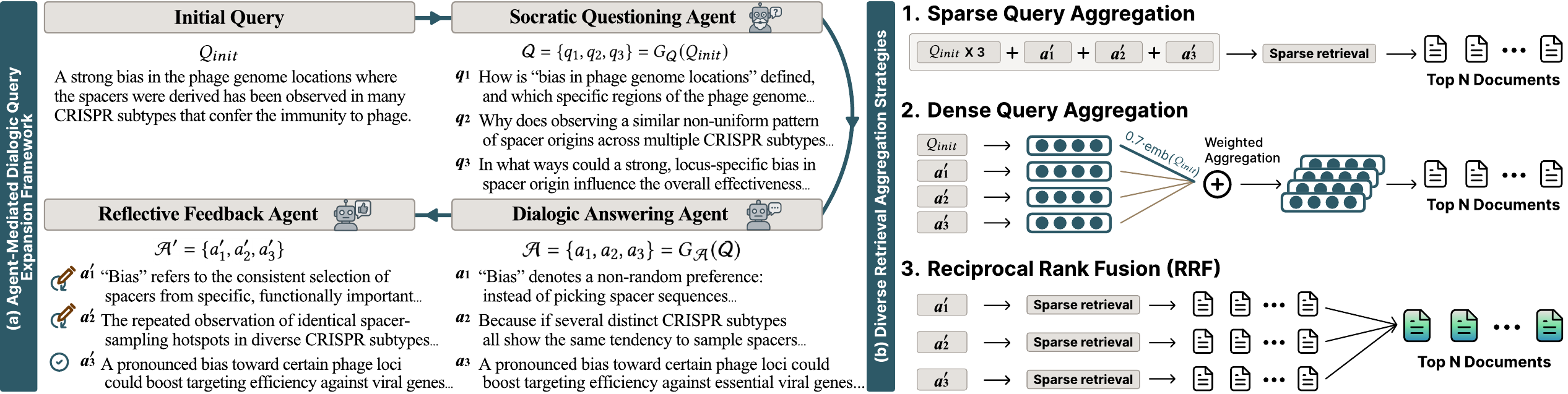}
    \caption{The overview of the proposed approach: (a) \underline{A}gent-\underline{M}ediated \underline{D}ialogic query expansion framework (left); (b) Retrieval aggregation strategies, where the enriched queries are utilized through 3  retrieval methods for enhanced information retrieval (right).}  
    \Description{The overview of the proposed framework. Given an initial query, the framework generates diverse relevant questions, produces corre- sponding pseudo-answers, and selectively rewrites and filters relevant answers.}
      \label{fig:framework}
\end{teaserfigure}

\received{20 February 2007}
\received[revised]{12 March 2009}
\received[accepted]{5 June 2009}

\maketitle

\section{Introduction}
Query expansion is a widely used technique in Information Retrieval (IR) for effectively improving search outcomes by enriching the initial query with additional contextual information~\cite{carpineto2012survey,azad2019novel,jagerman2023query}. Traditionally, methods such as Pseudo-Relevance Feedback (PRF) expand queries by selecting terms from top-ranked documents~\cite{robertson1990term,jones2006generating,lavrenko2017relevance}. While these conventional approaches have been successful to some extent, their reliance on static term selection limits the scope of expansion, leading to insufficient diversity and depth in capturing users’ information needs~\cite{roy2016using,imani2019deep}. 

In recent years, the emergence of Large Language Models (LLMs) has enabled more dynamic and sophisticated query rewriting techniques that leverage generative capabilities to address the limitations of traditional approaches~\cite{zhao2023survey,ye2023enhancing,liu2024query,lei2024corpus,seo2024gencrf,chen2024analyze}. Representative methods include \emph{Q2D}, which expands queries by generating pseudo-documents through few-shot prompting~\cite{wang2023query2doc}; \emph{Q2C}, which applies Chain-of-Thought (CoT) prompting to guide reformulation~\cite{jagerman2023query,wei2022chain}; \emph{GenQREnsemble}, which concatenates multiple sets of paraphrased, instruction-driven keywords with the initial query~\cite{dhole2024genqrensemble}; and \emph{GenQRFusion}, which fuses document rankings retrieved for each query-keyword combination~\cite{dhole2024generative}. However, these advances face notable challenges: (1) simplistic prompt variations often lead to homogeneous expansions lacking contextual breadth; (2) the absence of dynamic feedback mechanisms results in redundant or suboptimal content; and (3) the prevailing term or document level expansions generally lack structured reformulation, limiting their capacity to capture the full complexity of user intent.

To address these limitations, we introduce \textbf{AMD}: a novel \textbf{\underline{A}}gent-\textbf{\underline{M}}ediated \textbf{\underline{D}}ialogic framework that engages in a dialogic inquiry that leverages LLMs to perform query expansion. Inspired by recent advances demonstrating that Socratic questioning enhances language model reasoning and problem decomposition~\cite{liu2021dexperts,wu2023hence,qi2023art,ang2023socratic,goel2024socratic}, our process begins with a \emph{Socratic Questioning Agent} that reformulates the initial query into diverse, context-sensitive sub-questions, each targeting a specific Socratic questioning dimension~\cite{paul2019thinker} such as clarification, assumption, or implication to comprehensively explore the information need. These questions are then passed to a \emph{Dialogic Answering Agent}, which generates corresponding pseudo-answers that serve as knowledge units supplementing the initial query. Finally, a \emph{Reflective Feedback Agent} evaluates and further rewrites these pseudo-answers to ensure that only the most informative and relevant content is retained. This multi-agent process produces a multi-faceted augmentation of the initial query, enhancing its capacity to retrieve relevant information.

Extensive experiments on $8$ widely used benchmark datasets, including $6$ frequently used datasets from the BEIR Benchmark~\cite{thakur2021beir} and $2$ datasets from the TREC Deep Learning $2019$ and $2020$ tracks~\cite{craswell2020overview}, demonstrate that our \textbf{AMD} framework notably outperforms recent query expansion techniques across three retrieval methods. Overall, our contributions are: (1) a novel paradigm that leverages Socratic-inspired dialogic generation to reformulate the initial query and produce diverse, intent-aligned answers; (2) a dynamic feedback model that selectively rewrites generated answers for effective query augmentation; and (3) comprehensive empirical validation confirming that \textbf{AMD} generated expansions consistently enhance retrieval performance across diverse retrieval methods.

\section{Preliminaries}
\subsection{Query Expansion with LLM} Let \(Q_{init}\) denote the initial query and \(G\) be a Large Language Model (LLM) used for generation. Query expansion enhances retrieval by enriching initial query with additional context~\cite{azad2019query,claveau2020query,naseri2021ceqe,jia2023mill}. Two predominant LLM-based strategies have emerged: (1) \emph{Pseudo-Document Generation}, where \(G\) produces a surrogate document \(D\) or an expanded query \(Q^*\) to capture latent information~\cite{wang2023query2doc,jagerman2023query,zhang2024exploring}, and (2) \emph{Term-Level Expansion}, where \(G\) generates a set of terms \(T = \{t_1, t_2, \dots, t_M\}\) that reflect diverse aspects of \(Q_{init}\)~\cite{dhole2024genqrensemble,dhole2024generative,li2024can,nguyen2024exploiting}.

\subsection {Information Retrieval with Expanded Queries} In query expansion, Information Retrieval (IR) integrates the initial query with generated augmentations using various strategies. In detail, for \emph{sparse retrieval}, the common method concatenates multiple copies of the initial query with generated terms to reinforce core signals~\cite{wang2023query2doc,zhang2024exploring}. In \emph{dense retrieval}, one strategy directly combines the query and its expansions into an unified embedding~\cite{li2023pseudo,wang2023query2doc}, while another fuses separate embeddings from each component~\cite{seo2024gencrf,kostric2024surprisingly}. Additionally, \emph{Reciprocal Rank Fusion (RRF)} aggregates rankings from individual expanded queries by inversely weighting document ranks~\cite{mackie2023generative}. Our proposed \textbf{AMD} framework is compatible with all three retrieval methods, supporting flexible integration via sparse concatenation, dense embedding fusion, and RRF rank aggregation.

\section{Methodology}
Our proposed framework \textbf{AMD} operates through a dialogic interaction among $3$ specialized agents: (1) a \emph{Socratic Questioning Agent}, (2) a \emph{Dialogic Answering Agent}, and (3) a \emph{Reflective Feedback Agent}. An overview of the framework is illustrated in Figure~\ref{fig:framework}.

\subsection{Socratic Questioning Agent}
Prior studies have shown that structured Socratic questioning leads to more interpretable reasoning by prompting models to consider alternative perspectives, key concepts, and construct solutions~\cite{liu2021dexperts,wu2023hence,qi2023art,ang2023socratic,goel2024socratic}.
Motivated by these findings, we employ a \emph{Socratic Questioning Agent} to reformulate the user's information need into $3$ diverse and targeted sub-questions, each corresponding to a distinct Socratic questioning dimension (see Table~\ref{tab:subquestion-types}). Given an initial query \(Q_{init}\), this process is performed in a single LLM inference:
\begin{equation}\label{eq:question-generation}
\mathcal{Q} = \{q_1, q_2, q_3\} = G_\mathcal{Q}(Q_{init}),
\end{equation}
where \(G_\mathcal{Q}\) denotes the \emph{Socratic Questioning Agent}. Each sub-question \(q_i\) (for \(i = 1, 2, 3\)) is designed to reflect a specific Socratic perspective, thereby enriching intent understanding and promoting diversity.

\begin{table}[!htbp]
    \centering
    \begin{tabular}{lp{0.6\linewidth}}
    \toprule
        \textbf{Query Type} & \textbf{Role in Sub-question} \\
    \hline
        Clarification & Crafts a sub-question to refine intent and ensure accurate interpretation of the user query. \\
    \hline
    Assumption Probing & Decomposes the query by surfacing implicit assumptions, adding diversity and depth to retrieval. \\
    \hline
    Implication Probing & Explores downstream effects to expand the query with relevant and diverse information. \\
    \hline
\end{tabular}
\caption{Socratic-inspired sub-question types used for reformulating the initial query and generating diverse, intent-aligned expansions for enhanced information retrieval.}
\label{tab:subquestion-types}
\end{table}

\subsection{Dialogic Answering Agent}
For each sub-question \(q_i \in \mathcal{Q}\), the \emph{Dialogic Answering Agent} generates a corresponding pseudo-answer \(a_i\), serving as a surrogate document to enrich the query representation. This process reflects a dialogic exchange, where each answer is directly informed by the specific Socratic dimension and intent underlying its associated sub-question. All $3$ pseudo-answers are generated in parallel through a single LLM inference, and are formally defined as:
\begin{equation}\label{eq:answer-generation}
\mathcal{A} = \{a_1, a_2, a_3\} = G_\mathcal{A}(\mathcal{Q}),
\end{equation}
where \(G_\mathcal{A}\) denotes the \emph{Dialogic Answering Agent}. The resulting set \(\mathcal{A}\) offers a multi-perspective enrichment of the initial query, capturing diverse and intent-aligned aspects that improve retrieval robustness.

\subsection{Reflective Feedback Agent}
The \emph{Reflective Feedback Agent} critically assesses the complete set of three question-answer pairs \(\{(q_i, a_i)\}_{i=1}^3\) within the context of the initial query \(Q_{init}\). This agent applies reflective reasoning to evaluate and selectively rewrite each pseudo-answer, filtering out content that is vague, redundant, or irrelevant. This step aims to retain only the most informative and intent-aligned content, shaped by the prior Socratic and dialogic interactions, is retained. Importantly, our feedback agent is implemented using non-finetuned LLM, demonstrating the framework's generalizability and practicality. The refined output is formalized as:
\begin{equation}\label{eq:refined-answers}
\mathcal{A'} = G_\mathcal{F}(\{(q_i, a_i)\}_{i=1}^3, Q_{init}),
\end{equation}
where \(G_\mathcal{F}\) denotes the \emph{Reflective Feedback Agent}. The resulting set of refined pseudo-answers is:
\begin{equation}\label{eq:final-set}
\mathcal{A'} = \{a'_1, a'_2, a'_3\}.
\end{equation}
These $3$ refined answers \(\mathcal{A'}\) are then integrated with \(Q_{init}\) using various aggregation strategies, constructing an enriched query representation, enabling flexible adaptation to diverse retrieval techniques.

\begin{table*}[ht]
\centering
\renewcommand{\arraystretch}{1.3} 
\resizebox{\textwidth}{!}{%
\begin{tabular}{l|ccccccc|ccc|ccc}
\toprule
\multirow{3}{*}{\textbf{Methods}} & \multicolumn{7}{c|}{\textbf{BEIR Benchmark (nDCG@10)}} & \multicolumn{3}{c|}{\textbf{TREC DL'19}} & \multicolumn{3}{c}{\textbf{TREC DL'20}} \\
\cmidrule(lr){2-8} \cmidrule(lr){9-11} \cmidrule(lr){12-14}
 & Webis & SciFact & TREC-COVID & DBPedia & SciDocs & Fiqa & Avg. Score & nDCG@10 & R@1000 & Avg. Score & nDCG@10 & R@1000 & Avg. Score \\
\midrule
\multicolumn{14}{c}{\textbf{Sparse Retrieval Results}} \\
\hline
\emph{BM25} & 0.2719 & 0.6694 & 0.5868 & 0.2831 & 0.1592 & 0.2326 & 0.3672 & 0.4239 & 0.3993 & 0.4116 & 0.4758 & 0.4240 & 0.4500 \\
\emph{Q2C} (\citeyear{jagerman2023query}) & 0.3546 & \underline{0.6876} & 0.6954 & 0.3252 & \underline{0.1661} & 0.2595 & 0.4147 & 0.5439 & 0.4814 & 0.5127 & 0.5357 & 0.4941 & 0.5149 \\
\emph{Q2D} (\citeyear{wang2023query2doc}) & \underline{0.3679} & 0.6794 & \underline{0.6957} & \textbf{0.3378} & 0.1637 & \underline{0.2712} & \underline{0.4193} & \underline{0.5732} & \underline{0.4890} & \underline{0.5311} & \underline{0.5486} & \underline{0.4958} & \underline{0.5222} \\
\emph{GenQREnsemble} (\citeyear{dhole2024genqrensemble}) & 0.2887 & 0.5560 & 0.5104 & 0.2302 & 0.1058 & 0.2017 & 0.3155 & 0.4109 & 0.4110 & 0.4110 & 0.4261 & 0.4163 & 0.4207 \\
\emph{\textbf{AMD*}} (Sparse, Ours) & \textbf{0.3896*} & \textbf{0.7021*} & \textbf{0.7115*} & \underline{0.3352} & \textbf{0.1834*} & \textbf{0.2896*} & \textbf{0.4352*} & \textbf{0.5870*} & \textbf{0.4974*} & \textbf{0.5422*} & \textbf{0.5818*} & \textbf{0.5047*} & \textbf{0.5433*} \\
\midrule
\multicolumn{14}{c}{\textbf{Dense Retrieval Results}} \\
\hline
\emph{E5-Base} & 0.1786 & 0.6924 & 0.7098 & 0.4002 & 0.2326 & 0.3808 & 0.4324 & 0.7020 & 0.5185 & 0.6103 & 0.7029 & 0.5648 & 0.6339 \\
\emph{Q2C} (\citeyear{jagerman2023query}) & 0.1841 & 0.7028 & 0.7238 & \underline{0.4250} & 0.2595 & 0.4057 & 0.4502 & 0.5517 & 0.4891 & 0.5204 & \underline{0.7084} & 0.5715 & \underline{0.6400} \\
\emph{Q2D} (\citeyear{wang2023query2doc}) & \underline{0.1931} & \underline{0.7108} & \underline{0.7284} & 0.4229 & \underline{0.2712} & \underline{0.4094} & \underline{0.4560} & \textbf{0.7472} & \underline{0.5565} & \underline{0.6519} & 0.6971 & \underline{0.5799} & 0.6385 \\
\emph{\textbf{AMD*}} (Dense, Ours) & \textbf{0.1985} & \textbf{0.7324*} & \textbf{0.7493*} & \textbf{0.4435*} & \textbf{0.2871*} & \textbf{0.4135*} & \textbf{0.4707*} & \underline{0.7458} & \textbf{0.5752*} & \textbf{0.6605*} & \textbf{0.7128*} & \textbf{0.5847*} & \textbf{0.6488*} \\
\midrule
\multicolumn{14}{c}{\textbf{RRF Fusion (BM25-based) Results}} \\
\hline
\emph{GenQRFusion} (\citeyear{dhole2024generative}) & \textbf{0.3815} & \underline{0.6518} & \underline{0.6594} & \underline{0.2726} & \underline{0.1436} & \underline{0.2293} & \underline{0.3897} & \underline{0.4418} & \underline{0.4205} & \underline{0.4312} & \underline{0.4375} & \underline{0.4654} & \underline{0.4515} \\
\emph{\textbf{AMD*}} (RRF, Ours) & \underline{0.3749} & \textbf{0.6764*} & \textbf{0.6703*} & \textbf{0.3078*} & \textbf{0.1749*} & \textbf{0.2632*} & \textbf{0.4113*} & \textbf{0.5041*} & \textbf{0.4724*} & \textbf{0.4883*} & \textbf{0.5325*} & \textbf{0.4819*} & \textbf{0.5077*} \\
\bottomrule
\end{tabular}%
}
\caption{Combined retrieval performance on BEIR Benchmark (nDCG@10) and TREC DL'19/TREC DL'20 (nDCG@10 / R@1000). For BEIR, the Avg. column is the average across $6$ datasets. For TREC DL, the Avg. Score is computed as the average of nDCG@10 and R@1000. Bold indicates the best score and underline indicates the second-best score. * denotes significant improvements (paired t-test with Holm-Bonferroni correction, p $<$ 0.05) over the average baseline value for the metric.}
\label{tab:combined-results}
\end{table*}

\section{Experiments}
\subsection{Setup}

\paragraph{Datasets.} We evaluate our framework \textbf{AMD} on two widely used IR benchmark collections: (1) \emph{BEIR Benchmark}~\cite{thakur2021beir} and (2) \emph{TREC Deep Learning Passage Datasets}~\cite{craswell2020overview}. For \emph{BEIR Benchmark}, we specifically select $6$ widely used datasets: Webis-Touche2020~\cite{bondarenko2020overview}, SciFact~\cite{wadden2020fact}, Trec-COVID-BEIR~\cite{voorhees2021trec}, DBPedia-Entity~\cite{hasibi2017dbpedia}, SCIDOCS~\cite{cohan2020specter} and FIQA~\cite{maia201818}. For \emph{TREC Datasets}, we employ the Deep Learning Passage Tracks from $2019$ and $2020$, which consist of large-scale passage collections to ensure that our approach also performs well in challenging retrieval scenarios.


\paragraph{LLM and Retrieval Models.} For our \textbf{AMD} framework, we utilize an open-source model: \emph{Qwen2.5-7B-Instruct}\footnote{https://huggingface.co/Qwen/Qwen2.5-7B-Instruct}~\cite{qwen2.5}, which is a well-known high-performance LLM. We specifically use a temperature of $0.5$ and a maximum context length of $512$ during LLM inference. For the dense retrieval task, we employ \emph{multilingual-e5-base}\footnote{https://huggingface.co/intfloat/multilingual-e5-base}~\cite{wang2024multilingual} to encode both queries and documents into dense representations. The similarity between the encoded query and document embeddings is measured using cosine similarity, and top-ranked documents are retrieved accordingly. Additionally, we incorporate \emph{BM25}~\cite{robertson2009probabilistic} as a sparse retrieval baseline, specifically via using BM25s\footnote{https://github.com/xhluca/bm25s}~\cite{lu2024bm25s} library, a pure-Python implementation that leverages Scipy~\cite{virtanen2020scipy} sparse matrices for fast and efficient scoring.

\paragraph{Baselines and Our Approach.} 

We compare our \textbf{AMD} framework with state-of-the-art query expansion methods: (1) \emph{Q2D}~\cite{wang2023query2doc}: pseudo-document generation via few-shot prompting, (2) \emph{Q2C}~\cite{jagerman2023query}: Chain-of-Thought (CoT) guided reformulation, (3) \emph{GenQREnsemble}~\cite{dhole2024genqrensemble}: a concatenation of $10$ prompt-based keyword sets, and (4) \emph{GenQRFusion}~\cite{dhole2024generative}: an extended work of \emph{GenQREnsemble} that retrieves documents for each initial query paired with each set of keywords, and then fuses the resulting rankings to enhance retrieval performance. All baselines follow their original configurations. Additionally, our \textbf{AMD} framework enriches the initial query through a multi-agent dialogic process that generates $3$ distinct and targeted sub-questions, corresponding pseudo-answers, and refined feedback results, which promotes greater diversity in query expansion while ensuring both computational efficiency and alignment with user intent.

\subsection{Implementation Details}
\label{implementation-details}

\paragraph{Sparse Query Aggregation.}  
In the sparse retrieval setting, following previous work~\cite{wang2023query2doc,jagerman2023query}, we replicate the \(Q_{init}\) three times and then append all refined pseudo-answers \(a'_i\). Specifically, let \(Q_i = Q_{init}\) for \(i=1,2,3\). The expanded query is formulated as:
\begin{equation}
Q^*_{\text{sparse}} = \sum_{i=1}^{3} Q_{i} \oplus \sum_{j=1}^{|\mathcal{S}|} a'_j,
\end{equation}
where the "\(\oplus\)" operator denotes the concatenation of expanded  or reformulated queries, with [SEP] tokens as separators.

\paragraph{Dense Query Aggregation.}  
For dense retrieval, let \(\text{emb}(Q_{init})\) be the embedding of the initial query and \(\text{emb}(a'_i)\) the embedding of each refined pseudo-answer. Following previous work in weighted query aggregations~\cite{seo2024gencrf}, which employed a weight of \(0.7\) for the initial query embedding, we adopted the same weighting scheme and compute the final query embedding \(q^*\) as a weighted sum:
\begin{equation}
q^* = 0.7 \cdot \text{emb}(Q_{init}) + 0.3 \cdot \frac{1}{|\mathcal{S}|} \sum_{i=1}^{|\mathcal{S}|} \text{emb}(a'_i).
\end{equation}
\paragraph{Reciprocal Rank Fusion (RRF)}  
In the RRF setting~\cite{cormack2009reciprocal}, each refined pseudo-answer \(a'_i\) is used to form an individual expanded query \(Q^*_i\). For each document \(d\), let \(r_{i,d}\) denote its rank when retrieved with \(Q^*_i\). The final score for \(d\) is computed as:
\begin{equation}
\text{score}(d) = \sum_{i=1}^{|\mathcal{S}|} \frac{1}{k + r_{i,d}},
\end{equation}
where \(k = 60\) is a constant to dampen the influence of lower-ranked documents. Documents are then reranked based on their scores.

\subsection{Main Experiment Results}
In our experiments on both sparse and dense retrieval settings (see Table~\ref{tab:combined-results}), we found that methods such as \emph{GenQREnsemble}, \emph{Q2C}, and \emph{Q2D} offer incremental gains through query reformulations, yet each has notable limitations: \emph{GenQREnsemble} relies on $10$ diverse prompts for term-level expansion, resulting in redundant, narrowly focused outputs that miss the full spectrum of user intent, while also incurring high computational costs and lacking dense contextual understanding; \emph{Q2C} applies chain-of-thought (CoT) prompting but tends to generate repetitive and insufficiently diverse expansions; and \emph{Q2D} creates pseudo-documents that better reflect information needs, yet lacks effective filtering for less informative content. In contrast, our \textbf{AMD} framework reformulates each query into three Socratic sub-questions, generates corresponding pseudo-answers, and uses a feedback agent to dynamically evaluate and retain only the most informative expansions, ensuring multi-faceted, intent-aligned representations and notable improvements in retrieval performance. 

Similarly, in fusion-based retrieval, while methods such as \emph{GenQRFusion} aggregate rankings from up to $10$ separate LLM inferences, often introducing redundancy and inefficiency, our approach utilizes a targeted fusion strategy using filtered, high-quality expansions, minimizing computational overhead and consistently delivering superior results.

\subsection{Ablation Study and Analysis}

To assess the effectiveness of the \emph{Reflective Feedback Agent}, an ablation study was conducted on the BEIR Benchmark and TREC datasets. Table~\ref{tab:combined-avg} compares the retrieval performance of the full \textbf{AMD} framework against a variant without the feedback agent. The results indicate that incorporating the \emph{Reflective Feedback Agent} improves the overall average score, which showcases its ability to effectively filter out redundant and less informative pseudo-answers, ensuring that only high-quality expansions contribute to query augmentation.
\begin{table}[htbp]
    \centering
    \resizebox{\columnwidth}{!}{%
        \begin{tabular}{llccc}
            \toprule
            \textbf{Method} & \textbf{Feedback} & \textbf{BEIR} & \textbf{TREC DL'19} & \textbf{TREC DL'20} \\
            \midrule
            \multirow{2}{*}{Sparse Retrieval} 
                & (w/o) feedback & 0.4295 & 0.5353 & 0.5418 \\
                & (w) feedback   & \textbf{0.4352} & \textbf{0.5422} & \textbf{0.5433} \\
            \midrule
            \multirow{2}{*}{Dense Retrieval} 
                & (w/o) feedback & 0.4658 & 0.6523 & 0.6452 \\
                & (w) feedback   & \textbf{0.4707} & \textbf{0.6605} & \textbf{0.6488} \\
            \midrule
            \multirow{2}{*}{BM25/RRF Fusion}  
                & (w/o) feedback & \textbf{0.4147} & 0.4726 & 0.4985 \\
                & (w) feedback   & 0.4113 & \textbf{0.4883} & \textbf{0.5077} \\
            \bottomrule
        \end{tabular}
    }
    \caption{Combined average retrieval performance on \emph{BEIR Benchmark} and \emph{TREC DL} datasets with (w) and without (w/o) feedback. The scores are computed as the average over $6$ \emph{BEIR} datasets and for \emph{TREC DL'$19$} and \emph{TREC DL'$20$}. Bold values indicate the best performance for each method and metric.}
    \label{tab:combined-avg}
\end{table}

Moreover, the analysis reveals that our feedback agent not only boosts overall performance but also enhances robustness. In the absence of this component, performance variability increases and more noise from less relevant pseudo-answers is observed. In contrast, the refined feedback mechanism consistently maintains stable and superior retrieval effectiveness across diverse datasets. These findings underscore the importance of the \emph{Reflective Feedback Agent} in \textbf{AMD}, confirming dynamically rewriting high-quality expansions is crucial for capturing the multifaceted nature of user intent. Notably, even the framework without the evaluation module achieves a higher average score than the other baselines, which are shown in Table~\ref{tab:combined-results}.

\section{Conclusion}
In this paper, we presented \textbf{AMD}, a novel agent-mediated framework for query expansion that combines \emph{Socratic Questioning}, \emph{Dialogic Answering}, and \emph{Reflective Feedback Agents}. By reformulating queries into diverse sub-questions and iteratively refining pseudo-answers, \textbf{AMD} effectively produces robust, intent-aligned query representations. Extensive experiments on $8$ widely used benchmark datasets, including $6$ frequently used datasets from the BEIR Benchmark and $2$ datasets from the TREC Deep Learning $2019$ and $2020$ tracks confirm that our specially designed agent collaboration leads to consistent and significant improvements in retrieval performance.

\section{Limitations and Future Work}
While our \textbf{AMD} framework effectively improves query expansion via multi-agents, including (1) \emph{Socratic Questioning Agent}, (2) \emph{Dialogic Answering Agent}, and (3) \emph{Reflective Feedback Agent}, some residual noise may still persist in the expanded queries, as the current \emph{Reflective Feedback Agent} is based on a non-finetuned model and may not optimally filter out all irrelevant or redundant pseudo-answers. In future work, we plan to explore fine-tuning the feedback agent to further reduce such noise and further enhance selectivity. Additionally, more detailed qualitative analysis of each agent’s output, especially the impact of feedback rewriting, would offer deeper insights into their contributions.

\bibliographystyle{ACM-Reference-Format}
\bibliography{sample-base}

\appendix

\end{document}